\documentclass[a4paper,12pt]{article}
\usepackage{amsmath,amssymb,amsthm}
\usepackage{latexsym,graphicx,color,bbm,subfigure}
\usepackage{pdfsync}   
\usepackage{bbm}

\newcommand{\be}{\begin{equation}}
\newcommand{\ee}{\end{equation}}
\newcommand{\beq}{\begin{eqnarray}}
\newcommand{\eeq}{\end{eqnarray}}
\newcommand{\non}{\nonumber\\}

\newcommand{\p}{\partial}
\newcommand{\Tr}{{\rm Tr}}
\newcommand{\diag}{{\rm diag}}

\newcommand{\bea}{\begin{eqnarray}}
\newcommand{\eea}{\end{eqnarray}}
\def\Tr{ \hbox{\rm Tr}}
\def\tr{ \hbox{\rm tr}}

\def\bra{\langle}
\def\ket{\rangle}

\makeatletter \@addtoreset{equation}{section} \makeatother

\begin{document}

\thispagestyle{empty}
\begin{flushright}
IFUP-TH/2012-18
\end{flushright}
\vspace{10mm}
\begin{center}
{\Large   \bf Singular SQCD Vacua and Confinement} 
\\[15mm]
{Simone Giacomelli$^{a,b}$    and Kenichi Konishi$^{c,b}$  
} \footnote{\it e-mail address: si.giacomelli(at)sns.it, ~
konishi(at)df.unipi.it}
\vskip 6 mm

\bigskip\bigskip
{\it  
$^a$
  Scuola Normale Superiore, 
Piazza dei Cavalieri, 7, Pisa, Italy  
\\
$^b$
  INFN, Sezione di Pisa,
Largo Pontecorvo, 3, Ed. C, 56127 Pisa, Italy 
\\
$^c$  
~Department of Physics ``E. Fermi'', University of Pisa, \\
Largo Pontecorvo, 3, Ed. C, 56127 Pisa, Italy
\\
  }

\vskip 6 mm

\bigskip
\bigskip

{\bf Abstract}\\[5mm]
{\parbox{14cm}{\hspace{5mm}
\small

We revisit the study of confining vacua in the softly broken ${\cal N}=2$ supersymmetric QCD, in the light of some recent developments in our 
understanding  of the dynamics of ${\cal N}=2$ gauge theories. These vacua are characterized by an effective  magnetic $SU(r)$ gauge 
group ($0\leq r\leq N_f/2$) and are referred to sometimes as the $r$ vacua.  We further clarify the meaning of  $r  \leftrightarrow N_{f}-r$ duality arising from the matching of semi-classical  and  quantum vacua.   
A particular attention is paid  to certain singular SCFT's of ${\cal N}=2$ SQCD, driven into confinement phase by the adjoint mass deformation     
$\mu\, \Phi^{2}$.  In some cases they occur  as a result of coalescence  of different  $r$ vacua as the bare mass is tuned to a critical value. }
}

\end{center}
\newpage
\pagenumbering{arabic}
\setcounter{page}{1}
\setcounter{footnote}{0}
\renewcommand{\thefootnote}{\arabic{footnote}}

\tableofcontents

\section{Introduction} 

   A considerable progress is being made in our understanding of the dynamics of non-Abelian gauge theories in four dimensions.  A recent remarkable development concerns the better understanding of ${\cal N}=2$  superconformal theories (SCFT) \cite{AS}-\cite{CD}. Also many new results on the exact BPS spectra  in the strongly coupled gauge systems are now available (see e.g. \cite{CV1,GMN}).  Another venue in which considerable  development has occurred 
 is the investigations of soliton vortex and monopoles of non-Abelian type \cite{HT}-\cite{GJK}.   Together, it is quite plausible that these developments help clarifying many issues left still to be elucidated, even after the discovery of Seiberg-Witten solutions of ${\cal N}=2$ gauge theories and the developments which followed.  

One such important problem concerns the possible types of strongly-coupled gauge systems in confinement phase. It is the purpose of this paper to make a few remarks on
the confining vacua occurring in the softly broken ${\cal N}=2$  SQCD.  To fix an idea and for definiteness, we stick to the ${\cal N}=2$ supersymmetric QCD like theories with $SU(N)$ or $USp(2N)$ gauge theories with $N_{f}$ flavors of quark hypermultiplets, deformed by the adjoint mass perturbation, $\mu \, \Tr\, \Phi^{2}$.  

The paper will be  organized as follows. In Section 2 we discuss the simplest and well understood case of $SU(2)$ SQCD, describing the ideas that will play 
a crucial role in our analysis. In Section 3 we review the physics of the $r$ vacua and the mechanism of confinement which takes place when the 
$\mathcal{N}=1$ perturbation is turned on. In Section 4 we clarify the mechanism at the basis of the two-to-one map relating semiclassical and 
quantum vacua observed in \cite{CKM}. This correspondence links $r$ and $N_f-r$ vacua and is reminiscent of Seiberg's duality. In Section 5 we discuss 
the low energy physics at a large class of fixed points in $\mathcal{N}=2$ SQCD. These SCFTs become confining when we turn on the $\mathcal{N}=1$ 
perturbation and we discuss the mechanism of confinement  occurring in these special cases. We conclude with a discussion in Section 6.

\section{Softly broken ${\cal N}=2$ $SU(2)$ SQCD}   

The softly broken  ${\cal N}=2$  $SU(2)$ gauge theory with $N_{f}=1,2,3$  quark hypermultiplets has extensively been studied starting from the pioneering work by Seiberg and Witten \cite{SW1}-\cite{SUN}.  In order to introduce some of the issues 
discussed later let us briefly review the physics of the $N_{f}=2$ theory.   
There are four singularities in the $u=
\bra  \Tr \Phi^{2} \ket$ plane where some hypermultiplets become massless. The effective theory at these points is Abelian, dual $U(1)$ gauge theory. 
For small, nearly equal bare quark masses,  $m_{1}\simeq m_{2} \ll \Lambda$,     the singularities group into two pairs of nearby  singularities. 
The massless hypermultiplets in these two singularities are the Abelian monopoles in one or the other of the spinor representations 
\beq   ({\underline 2}, {\underline 1})\qquad   {\rm  or} \qquad   ({\underline 1}, {\underline 2})   \label{JR} \eeq 
of the flavor symmetry group $SO(4)\sim SU(2)\times SU(2)$.  

When the perturbation  $\mu \, \Tr \Phi^{2}$  ($\mu \ll \Lambda$)   is added in the system,  the monopole, say in  $({\underline 2}, {\underline 1})$, condenses, 
\beq   \bra M_{1} \ket \sim \sqrt{\mu \Lambda}\;,     \label{mpcondense}
\eeq
 the dual $U_{D}(1)$ gauge group is Higgsed, and the system is in confinement phase. 
An interesting feature of this case is that the confinement order parameter at the same time breaks the global symmetry as  
\beq    SU(2)\times SU(2) \to SU(2)\;. \label{symbrk}
\eeq
 The Seiberg-Witten effective action correctly describes the low-energy excitations:  the exactly massless Nambu-Golstone bosons of the symmetry breaking  (\ref{symbrk})  and their superpartners.  Unlike the light flavored standard  QCD,  the massless Nambu-Goldstone bosons do not carry  the quantum numbers of the remaining unbroken $SU(2)$.
There are also light but massive dual photon and dual photino of the order of $\sqrt{\mu \Lambda}$, which arise as a result of the dual Higgs mechanism.  

 All these light particles are gauge invariant states (they are asymptotic states); the presence of the original quarks degrees of freedom can be detected in the flavor quantum numbers \cite{JR}.  
 
The low energy system is a dual Abelian $U(1)$ gauge theory broken by the magnetic monopole condensation (\ref{mpcondense}).   The ANO vortex of this system, with tension $\sim \mu \Lambda$,  carries the (Abelianized) chromoelectric flux.  The fact that the underlying  $SU(2)$ theory is simply connected, means that such a vortex must end:  the endpoints are the quarks (and squarks) of the underlying theory.  Quarks are confined. 

An important point we want to stress is the fact that the particles becoming massless at each Abelian singularity of the Seiberg-Witten curve  are  pure magnetic monopoles even though they carry distinct labels  $\{n_{m\,i}, n_{e\, i}\}$ ($i=1,2, \ldots, N-1$)   and coupled to different ``magnetic duals'',  $n_{m} A_{D \, \mu} + n_{e} A_{\mu}$. 
    In the case of $SU(2)$ theory where  there is only $U(1)$ gauge interactions at low energies  this fact is easily seen \cite{KT}.
 At a singularity of the
quantum moduli space
where the  $(n_m, n_e)$ dyon   becomes massless  
\be n_m   a_D  +  n_e   a =0, 
\ee
the exact SW  solution tells us that 
\beq 
\frac{n_m (d a_D / du) +  n_e (d a / du) }{ (d a / du)}  =0, 
\label{theta0}
\eeq
due to a logarithmic singularity in the denominator.  Thus
\beq 
\theta_{eff} = \Re \frac{d a_D }{ da}  \, \pi = - \frac{n_e }{ n_m}\pi
\label{theeff}
\eeq 
and  the electric charge of  a $(n_m, n_e)$ ``dyon''  is   \cite{Witten}
 \beq
\frac{2 }{ g}Q_e = n_e + \frac{\theta_{eff} }{ \pi} n_m =0.
\label{dyonismonop}
\eeq

In the case of the $SU(2)$ theory with $N_{f}=2$ with small masses, the massless dyons (which condense upon $\mu \Phi^{2}$ perturbation)  carry $(n_{m},n_{e})$
charges  
\be  (n_{m},n_{e}) = (1,0), \qquad  \theta_{eff} =0\;, 
\ee
in one doublet of singularities,  and 
\be  (n_{m},n_{e}) = (1,1), \qquad  \theta_{eff} = -  \pi\;,  
\ee
in the other.   Thus in all vacua  the quarks (with charges $(n_{m}^{1},n_{e}^{1})= (0,1)$) are confined, carrying a relative nonzero  Dirac unit 
\be   D = n_{m}^{1}  n_{e}^{2} -     n_{m}^{2}  n_{e}^{1} \qquad Mod \,\,{2},  
\ee
with respect to the condensed fields, $(n_{m}^{2}, n_{e}^{2})$.  In the $m \to 0$  limit, a ${\mathbbm Z}_{2}$ symmetry ensures that the physics at the two vacua look identical,  even though the light monopoles (dyons) are coupled locally to two 
different magnetic duals.

As was shown in \cite{SW2}, all massless ``dyons'' in $SU(2)$ theory with various $N_{f}$  carry  $n_{m}=1$.  Their condensation upon the $\mu \Phi^{2}$  perturbation leads to quark confinement. The only 
 exception occurs \cite{SW2}  in one of the vacua  of $N_{f}=3$ theory,  where  massless $(2,1)$ dyons appear as the infrared degrees of freedom.  This vacuum  (where $\theta_{eff}= - 1/2$)  survives the $\mu \Phi^{2}$ deformation,  the $(2,1)$ dyons condense,  but  quarks are unconfined:  it is in  't Hooft's  oblique confinement phase \cite{TH}.  The phase  of the  pure (non supersymmetric)  $SU(2)$ Yang-Mills theory with $\theta = -\pi$ is believed to be in such a phase, where the composite of the $(1,0)$
 monopole and the $(1,1)$ dyon with charges $\mp\tfrac{1}{2}$ condenses.

\section{The quantum $r$ vacua \label{quantumr}}

The classical and quantum moduli space of the vacua of the ${\cal N}=2$ supersymmetric $SU(N)$ QCD has been first studied systematically by Argyres, Plesser and Seiberg and others \cite{APS,Others,CKM}.  
 Of particular interest are the $r$-vacua characterized by an effective low-energy $SU(r)\times U(1)^{N-r}$ gauge symmetry, with massless monopoles carrying the 
 charges shown in  the Table~\ref{tabnonb}  (taken from ~\cite{APS}).   When the adjoint scalar mass  $\mu \Tr\Phi^{2}$  term, which breaks supersymmetry to  ${\cal N}=1$,
is added the massless Abelian  ($M_{k}$)  and non-Abelian monopoles ($\cal M$) all condense, bringing the system to a confinement phase.
\begin{table}[b]
\begin{center}
\vskip .3cm
\begin{tabular}{|ccccccc|}
\hline
&   $SU(r)  $     &     $U(1)_0$    &      $ U(1)_1$
&     $\ldots $      &   $U(1)_{N-r-1}$    &  $ U(1)_B  $  \\
\hline
$n_f \times  {\cal M}$     &    ${\underline {\bf r}} $    &     $1$
&     $0$
&      $\ldots$      &     $0$             &    $0$      \\ \hline
$M_1$                 & ${\underline {\bf 1} } $       &    0
&
1      & \ldots             &  $0$                   &  $0$  \\ \hline
$\vdots $  &    $\vdots   $         &   $\vdots   $        &    $\vdots   $
&             $\ddots $     &     $\vdots   $        &     $\vdots   $
\\ \hline
$M_{N-r-1} $    &  ${\underline {\bf 1}} $    & 0                     & 0
&      $ \ldots  $            & 1                 &  0 \\ \hline
\end{tabular}
\caption{The massless non-Abelian and Abelian monopoles  and their charges  at the $r$ vacua
at the root of a ``non-baryonic'' $r$-th   Higgs branch.  }
\label{tabnonb}
% \vskip .3cm
\end{center}
\end{table}
The form of the effective action describing these light degrees of freedom is dictated by the ${\cal N}=2$ supersymmetry and the gauge and flavor symmetries. 
The effective superpotential has the form ~\cite{APS,Others,CKM}
   \bea
W_{r-vacua} &=& \sqrt2 \,  \Tr ({\cal M} \phi {\tilde {\cal M}}) + \sqrt2  \, a_{D 0} \Tr ({\cal M}
{\tilde {\cal M}}) + \sqrt 2 \sum_{k=1}^{N-r-1} a_{D k} M_k {\tilde M}_k + \nonumber \\
&+& \mu \left(\Lambda \sum_{k=0}^{N-r-1} \,c_k  a_{D  k} + \frac{1}{ 2} \Tr \phi^2\right),
\label{nonbaryonic}
\eea
where $\phi$ and $a_{D\, 0}$ are the adjoint scalar fields in the  ${\cal N}=2$    $SU(r)\times U(1)$ vector multiplet, $a_{D\,  k}$, $k=1,2,\ldots, N-r-1$  are the adjoint scalars of the Abelian $U(1)^{N-r-1}$ gauge multiplets. $M_{k}$'s are  the Abelian monopoles, each carrying one of the magnetic $U(1)$ charges, whereas ${\cal M}$ (with $r$ color components and in the fundamental representation of the flavor $SU(N_{f})$ group)  are the non-Abelian monopoles.   The terms linear in $\mu$ is generated by the microscopic   ${\cal N}=1$ perturbation $\mu \, \Tr \Phi^{2}$, written in terms of the infrared degrees of freedom  $a_{D k}$ and $\phi$, and $c_{k}$ are some dimensionless  constants of order of unity.    These  quantum $r$-vacua are known to exist only for $r \le  \left[ \frac{N_{f}}{2} \right]$. 
 
When small, generic bare quark mass terms 
\beq     W_{masses} =  m_{i} \, Q_{i} {\tilde Q}_{i}
\eeq
are added in the microscopic theory,
 the infrared theory gets modified further by the addition 
\beq
\Delta W_{masses} = m_i  \, {\cal M}^{i} {\tilde {\cal M}}_i + \sum_{k=1}^{N-r-1} S_k^j\,
m_j  M_k {\tilde M}_k ,\qquad  (i,j =1,2,\ldots, N_{f}),
\label{masses}\eeq
where $S_{k}^{j}$ are the $j$-th quark number carried by the $k$-th monopole.  
Supersymmetric vacua are found by minimizing the potential following from Eq.~(\ref{nonbaryonic}) with Eq.~(\ref{masses}),  
and by vanishing of the $D$-term potential.  

   The part of the decoupled $U(1)^{N-r-1}$ theory involving  Abelian monopoles  is trivial and gives the VEV's 
   \beq   a_{D k}  \sim O(m_i); \quad    M_k= {\tilde M}_k \sim \sqrt{\mu
\Lambda}\;,   \label{asinsu2} \eeq
as in the $SU(2)$  theories.

   The  equations for the $SU(r)\times U(1)$ sector   (see Eq.~(\ref{effD1})-Eq.~(\ref{effF4})) are less trivial.  The equations look rather similar to the  
semiclassical equations of the microscopic $SU(N)$ theory,  which are valid for $\mu \gg \Lambda$,  $|m_{i}| \gg \Lambda$  
(Appendix A),  but there are a few crucial differences. 

One is that the effective gauge group $SU(r)\times U(1)$ is not simply connected and the low-energy system generates vortex solutions, while the microscopic theory cannot possess
such solitons.   Secondly,  the massless hypermultiplets in the system describe magnetically charged particles,  in contrast to those in the  
original ultraviolet Lagrangian.  Finally,  the range of validity of the effective theory is limited to the excitations of energies much less than the dynamical scale $\Lambda$, as the particles of masses of the order of $\Lambda$ or larger have been integrated out in obtaining it.

This last fact makes  the identification of the correct solutions of  Eq.~(\ref{effD1})-Eq.~(\ref{effF4}) somewhat a subtle task
(i.e., fake solutions involving VEVs of the order of $\Lambda$ must be disregarded):
the solutions are given by \cite{CKM}:  
\beq   \phi =\frac{1 }{ \sqrt{2}} \, \left(\begin{array}{ccc}-m_1- \sqrt{2}  \psi_0 &  &  \\ & \ddots &  \\ &  & -m_r- \sqrt{2} \psi_0\end{array}\right)\;, 
    \label{phiveveff}   \eeq
\beq  {\cal M}_a^i  =
\left(\begin{array}{cccc}d_1 &  &  &  \\    & \ddots &    &  \\  &  & d_r &      \\   &  &    & {\mathbf 0}\end{array}\right)\;, \qquad {\tilde {\cal M}}_i^a  =
\left(\begin{array}{cccc}{\tilde d}_1 &  &  &  \\    & \ddots &    &  \\  &  & {\tilde d}_r &      \\   &  &    & {\mathbf 0}\end{array}\right)\;, 
\eeq
where   $d_i$, ${\tilde d}_i$'s   and $\psi_{0}$  are given by
\beq  \psi_0= -\frac{  1 }{ \sqrt{2} \,   r}    \sum_i m_i , \label{corrvacua}  \eeq
\beq    d_i  {\tilde d}_i  = \mu \left( m_i - \frac{1}{r}\sum_j^r
m_j\right) - \frac{\mu
\Lambda}{\sqrt{2}\,  r}\;.  \eeq
%\beq  M_k {\tilde M}_k  =  -\mu \Lambda,  \eeq
In the limit $m_{i}\to 0$   the monopole VEV's tend to 
\beq     \bra  {\cal M}_{a}^{i}     \ket   =\delta_{a}^{i} \, \sqrt{\frac{\mu
\Lambda}{\sqrt{2}\,  r}}, \quad i, a =1,2,\ldots, r, \qquad    \bra  {\cal M}_{a}^{i}     \ket  =0,  \quad  i = r+1, \ldots, N_{f}\;.   \label{cflock}
\eeq
The system is in a color-flavor locked phase of the dual $SU(r)$ gauge theory. The flavor $SU(N_{f})\times U(1)$  symmetry of the underlying SQCD is 
dynamically broken as
\beq       SU(N_{f})\times U(1) \to  U(r) \times U(N_{f}-r)\;.  \label{symbr}
\eeq
The fact that $ \bra  {\cal M}_{a}^{i}     \ket $  is nonvanishing in the limit $m_{i}\to 0$ means that the symmetry breaking is dynamical, 
and this property distinguishes the $r$ vacua appearing at these nonbaryonic roots from the vacua at the baryonic root.   \footnote{ The vacua at the baryonic root, present only for $N_{f}> N$,  are interesting as they are characterized \cite{APS} by the low-energy effective $SU({\tilde N})$ gauge group, ${\tilde N}\equiv N_{f}-N$. 
It was indeed argued \cite{APS} that these might be relevant for the  understanding of the Seiberg duality in the ${\cal N}=1$ SQCD, and some further observations on this point were made recently \cite{SYrec}.  These vacua at the baryonic root are however nonconfining \cite{CKM} in the limit $m \to 0$, $\mu \ne 0$.
 } 

Until now we have discussed only the case of degenerate (or slightly unequal) bare masses $m_i$ for the flavors. An important observation is the 
fact that, when the $m_i$'s are generic, each $r$ vacuum splits in $\binom{N_f}{r}$ Abelian vacua. Taking into account the Witten effect as in Section 2, 
it is easy to generalize the argument we gave for $SU(2)$ to $SU(N)$ and conclude, as we will now see, that the particles becoming massless in each one of these vacua 
are magnetic monopoles and not dyons.
Consider a Cartan basis for $SU(N)$: 
 \begin{equation}   [H_i, H_k]=0, \qquad (i,k=1,2,\ldots, r);  
    \qquad   [H_i,  E_{\alpha}] = \alpha_i \, E_{\alpha}; \qquad
    [E_{\alpha}, E_{-\alpha}]= \alpha^i  \, H_i; 
  \end{equation}
  \begin{equation}   [E_{\alpha}, E_{\beta} ]=  N_{\alpha \beta}\,
    E_{\alpha + \beta} \qquad (\alpha+ \beta \ne 0).  
  \end{equation}
where $\alpha$'s are the root vectors.    $3(N-1)$  generators can be grouped into  $SU(2)$ subsets of  generators,  
\beq \qquad   [H_i,  E_{\alpha}] = \alpha_i \, E_{\alpha}; \qquad
    [E_{\alpha}, E_{-\alpha}]= \alpha^i  \, H_i,
    \eeq
containing $N-1$ diagonal $U(1)$ generators. 

Assuming  Abelianization  the magnetic monopoles are the 't Hooft-Polyakov monopoles living in these broken $SU(2)$ groups.  Each of the $SU(2)$ group acquires a $\theta$ term,
\beq   \frac{\theta}{32\pi^{2}}    \sum_{j=1}^{3}  F_{\mu \nu}^{j} \tilde F^{j \, \mu \nu}  
=  \frac{\theta}{8\pi^{2}}      \sum_{j=1}^{3}   {\bf E}^{j} \cdot   {\bf B}^{j},
\eeq
The $i$-th magnetic monopole contributes to the electromagnetic static energy
\beq   \frac{\theta}{8\pi^{2}}      \sum_{j=1}^{3}   {\bf E}^{j} \cdot   {\bf B}^{j}=\
\frac{\theta}{8\pi^{2}}    (-\nabla \phi) \cdot  \nabla \frac{  g_{m}}{r} =\frac{\theta}{8\pi^{2}}  \phi \,  \nabla^{2} \frac{  g_{m}}{r} =-\frac{\theta}{2\pi} g_{m} \phi\, \delta^{3}({\bf r})\;. 
\eeq
thus carries the $i$-th ``electric''  $U(1)$ charge,  $-\frac{\theta}{2\pi} g_{m}$.  Of course,  the monopole of the $i$-th  $SU(2)/U(1)$ sector  ($i=1,2,\ldots N-1$) is neutral with respect to all other $U(1)$'s. 

Under the dynamical hypothesis of Abelianization,  thus each $U(1)$ factor has its own Witten effect.  The argument made in the $SU(2)\to U(1)$ theories   works here too. 

In the equal mass limit,  the $r$ vacua with their nonAbelian sector are recovered. The massless and light particles in various Abelian vacua now 
(nontrivially)  recombine into multiplets  of the $SU(r)$ gauge group. We  conclude that these are nonAbelian monopoles and dual gauge fields.

\subsection{Low-energy excitations and nonAbelian chromoelectric vortices}

The obvious low-energy excitations of this system are the massless and light particles described by the effective Lagrangian described above. These can be found by expanding around the vacua (\ref{phiveveff})-(\ref{cflock}).  They contain massless Nambu-Goldstone bosons of the breaking (\ref{symbr}) and their superpartners, as well as light pseudo Nambu-Goldstone particles of the $SU_{R}(2)$  breaking. Also, the dual $SU(r)\times U(1)^{N-r}$ gauge bosons and gauginos form light massive multiplets. 

What is perhaps not so well known (however, see \cite{SYrec} for related remarks)  is the fact that, apart from these elementary excitations, the system described by Eqs.(\ref{phiveveff}) and (\ref{masses})  has low-energy nonAbelian excitations of a different sort.  As $\Pi_{1}(U(r) \times U(1)^{N-r-1}) = {\mathbbm Z}^{N-r}$, the low-energy system possesses soliton vortices. In the vacuum (\ref{cflock})  the minimum vortex configuration  (see Eq.~(\ref{minivort1})) breaks the color-flavor diagonal symmetry to $SU(r-1)\times U(1)$: it is a nonAbelian vortex \cite{HT}-\cite{GJK}.  The fluctuation of the orientational modes of 
\be  CP^{r-1}= SU(r)/SU(r-1)\times U(1)
\ee
  is described by a vortex worldsheet sigma model, 
\begin{align}
S_{1+1}   & =   2\beta \int dtdz \; \tr\left\{
X^{-1}\p_\alpha B^\dag
Y^{-1}\p_\alpha B
\right\}   \non
&=  2\beta \int dtdz \; \tr\left\{
\left(1 + B^\dag B\right)^{-1}\p_\alpha B^\dag
\left(\mathbf{1}_{r-1} + B B^\dag\right)^{-1}\p_\alpha B
\right\} \ , \label{eq:sigmamodelaction}
\end{align}
where $B$, a $r-1$ component vector,  represents the inhomogeneous  coordinates of $CP^{r-1}$  (see Eq.~(\ref{eq:Umatrix})) and   $\beta$ is a constant. 
The low energy system has also $N-r-1$  distinct Abelian (Abrikosov-Nielsen-Olesen) vortices, as the dual $U(1)^{N-r-1}$  theory is in the Higgs phase (see Eq.~(\ref{asinsu2})). 

The point of crucial importance is the fact that the underlying $SU(N)$ theory, being simply connected, 
does not support a  vortex solution.  It means that  both the nonAbelian vortex (\ref{eq:sigmamodelaction}) and the Abelian vortices of the $U(1)^{N-r-1}$  sectors  must end. 
These vortices in the dual, magnetic theory  carry chromoelectric fluxes.  The endpoints are quarks of the fundamental theory,  which, 
being relatively nonlocal to the low-energy effective degrees of freedom, and also  having dynamical masses of the order of $\Lambda$, 
are not explicitly visible in the low-energy effective action \footnote{Not all effects related to the underlying quarks are invisible at low energies, however. The zero-energy quark fermion modes are indeed  responsible for giving the flavor quantum numbers to the monopoles \cite{JR}
as  in Table~\ref{tabnonb}.}.  The quarks are confined.\footnote{Of course, as the underlying theory contains scalars in the fundamental representation there are no distinct 
phases between the confinement and Higgs phase in these theories (complementarity).} 

The system produces more than  one kinds of confining strings as the $SU(N)$ gauge symmetry of the ultraviolet theory is dynamically broken to $U(r)\times U(1)^{N-r-1}$ at low energies;  the mesons appear in various Regge trajectories of different slopes. The only exception is the case of $SU(3)$ theory,
where  the only nontrivial $r$ vacua ($r=2$) corresponds to a low energy $U(2)$ theory.  Since 
$  \Pi_{1}(U(2)) = {\mathbbm Z}$  there is a unique universal Regge trajectory.

\section{$r   \leftrightarrow  N_{f}-r$  duality}

One of the remarkable facts noted in \cite{CKM}, \cite{BKM} (see also more recent observations \cite{Simone,SYrec} on this point)  is the fact that semiclassical  $r$  vacua  (defined at large and equal quark masses $m_{i}$)  are related  nontrivially  to quantum $r$ vacua.   First of all, while there are semi-classical $r$ vacua with $r=0,1,2,\ldots, {\rm Min}\, \{N_{c}-1, N_{f}\}$  there are quantum $r$
vacua only up to  $r= N_{f}/2$.  This latter fact can be understood as due to the renormalization group behavior of the magnetic monopoles: 
the dual $SU(r)$  group is only infrared-free as long as $r < N_{f}/2$.  
 In Ref.\cite{CKM,BKM}  the two-to-one correspondence (both $r$ and $N_{f}-r$  classical vacua flow into  
quantum $r$ vacua)  was suggested by the counting of the number of vacua having  the same global symmetries.

%The vacua of $\mathcal{N}=2$ SQCD with $SU(N)$ gauge group which are not lifted by adding a mass term for the adjoint field, are 
%labelled by an integer $r$. Both the pattern of flavor symmetry breaking and the counting of vacua suggest the correspondence 
%$r, N_f-r\rightarrow r$ between classical and quantum vacua respectively. The analysis of \cite{APS,CKM} show that the low energy 
%physics at quantum r vacua is characterized by a nonAbelian $SU(r)$ gauge symmetry with $N_f$ massless matter fields in the fundamental 
%representation. 

Such a correspondence implies that the low energy physics of the classical  $r$ and $N_f-r$  vacua is the 
same. To investigate this point, we need somehow a nonperturbative definition of classical $r$ vacua and this was achieved in 
\cite{noi}. Each vacuum can be obtained placing $N_f$ poles on the $\mathcal{N}=1$ curve at $z=-m/\sqrt{2}$ ; some on the first 
sheet and the others on the second one. If we take the large $m$ limit, it is easy to see that vacua characterized by $r$ poles 
on the first sheet match precisely with classical $r$ vacua (this is what we mean by non perturbative definition).

In the semiclassical regime (for $m\gg\Lambda$), classical $r$ vacua 
with $r<\frac{N_f}{2}$ are characterized by a nonAbelian $SU(r)$ gauge group which is infrared free: the Higgs mechanism that breaks $SU(N)$ 
to $SU(r)\times U(1)^{N-r-1}$ occurs at high energies ($\sim m$), where the coupling costant is very small. The theory remains 
at weak coupling at all energy scales and we do not expect quantum corrections to change this picture dramatically. As we let 
$m$ decrease, quantum corrections will become relevant at energy scales of order $\sim\Lambda$, but the physics in the infrared 
will not be modified. In the massless limit we then obtain precisely one of the quantum $r$ vacua described in \cite{APS,CKM}.

If we instead consider a classical $r$ vacuum with $r>\frac{N_f}{2}$, from the equations of motion alone we cannot say much 
about the behavior at low energies: the theory at energy scales below $m$ is now asymptotically free and the coupling 
constant grows in the infrared. Our claim is that this theory admits in the infrared a weakly coupled dual description equivalent 
to the low energy effective action for $r'= N_f-r <\frac{N_f}{2} $ vacua. In order to prove this, we must show that the SW curve factorizes in the 
same way for $r$ and $N_f-r$ classical vacua, as we will now explain. In \cite{APS} it was shown that the vacua which are not lifted 
by the $\mathcal{N}=1$ perturbation $\mu\Tr\Phi^2$ are all the points in the moduli space such that the SW curve factorizes as 
\beq\label{fac} y^2=(x+m)^{2r}Q_{N-r-1}^{2}(x)(x-\alpha)(x-\beta),\quad r\leq\frac{N_f}{2},\eeq and in such a vacuum the effective low energy theory 
includes an Abelian $U(1)^{N-r-1}$ sector and a nonAbelian one which is an infrared free $U(r)$ gauge theory with $N_f$ 
massless flavors. From the above discussion it is clear that classical $r$ vacua with $r<N_f/2$ fall in this class. In order 
to prove our claim, we thus need to show that the SW curve factorizes as above both for classical $r$ and $N_f-r$ vacua.

In order to determine the form of the SW curve at classical $r$ vacua, our starting point will be the equation relating the SW 
curve and the chiral condensates \cite{CDSW}:
\be P_{N}(z)=z^{N}e^{-\sum_{i}\frac{U_i}{z^i}}|_{+}  +\Lambda^{2N-N_f}\frac{(z+m)^{N_f}}{z^N}e^{\sum_{i}\frac{U_i}{z^i}}|_{+},  \label{SWcurve}  \ee
where $U_i$ are the vacuum expectation values  $U_{i}=\frac{1}{i}\langle\Tr\Phi^i\rangle$ and  the symbol $ ...|_{+} $ indicates that only terms with nonnegative powers of
$z$ are kept ($P_{N}(z)$ is thus a polynomial).    These in turn can be computed from the generalized Konishi anomaly relations \cite{CDSW}  adapted to 
the case of  $SU(N)$  SQCD with the  ${\cal N}=1 $ adjoint scalar mass term  $\mu \, \Tr \Phi^{2}$ \cite{noi}\;:  
\be \left\langle\Tr\frac{1}{z-\Phi}\right\rangle=\sum_{i\geq0}\frac{\langle\Tr\Phi^i\rangle}{z^{i+1}}=
\frac{\frac{N_f-2r}{2}\sqrt{\mu^2(a+m)^2-4S\mu}}{(z+m)\sqrt{\mu^2(z-a)^2-4S\mu}} + \frac{N_f/2}{z+m}+
\frac{\mu(N-N_f/2)}{\sqrt{\mu^2(z-a)^2-4S\mu}},   \label{KArel}   \ee
where $S$ and  $a$ are the gaugino and meson condensates 
$$    S  \equiv  \frac{g^{2}}{32 \pi^{2}} \langle W^{\alpha}  W_{\alpha}  \rangle\;;
\qquad       a \equiv \frac{\sqrt{2}}{N\mu}\langle\tilde{Q}^{i}Q_{i}\rangle\;. $$
These can be determined from the Dijkgraaf-Vafa superpotential. The corresponding equations, determining at once $S$ and $a$ both 
for $r$ and $N_f-r$ classical vacua, are \cite{noi}
\footnote{As explained in \cite{noi}, the equations obtained by extremizing the DV superpotential can be brought in this form 
only for $r\geq N_f-N$. This will be enough for our purpose of discussing the $r\leftrightarrow N_f-r$ correspondence. 
The equations in \cite{noi} look  slightly different from the ones given here. This is due to a different normalization used:  if $m'$ and $\Lambda'$ denote the 
parameters used in \cite{noi}  the following relations hold: $m=m'/\sqrt{2}$ and $\Lambda^{2N-N_f}=\sqrt{2}^{N_f}(\Lambda')^{2N-N_f}$.}
\be \left(\frac{N-r}{N_f-2r}a-\frac{r}{N_f-2r}m\right)^{N-r}\left(\frac{N_f-N-r}{N_f-2r}a+\frac{N_f-r}{N_f-2r}m\right)^{N+r-N_f}=\Lambda^{2N-N_f}, \label{eqxa}  \ee
\be  S=\mu\left(\frac{N-r}{N_f-2r}a-\frac{r}{N_f-2r}m\right)\left(\frac{N_f-N-r}{N_f-2r}a+\frac{N_f-r}{N_f-2r}m\right). \label{eqxS}   \ee
Of the $2N-N_f$ solutions $N-r$ describe classical $r$ vacua and the remaining $N+r-N_f$ correspond to $N_f-r$ vacua. We can 
distinguish the two groups of solutions exploiting the fact that $S$ tends to infinity in the large $m$ limit only for classical $r$ 
vacua with $r<N_f/2$ \cite{noi}.
%As they determine the SW curve both for $r$ and $N_f-r$ vacua through  (\ref{SWcurve})  its factorization 
%property in the (classical) $r$ and $N_{f}-r$ vacua are the same and the infrared physics is described in both cases by the quantum system discussed in Section \ref{quantumr}. 

Solving these equations in general is very hard and we will not attempt to do so. However, in the massless limit they greatly 
simplify making it possible to check our claim.
In the massless case the equation for $a$ can be easily solved, leading to the $2N-N_f$ solutions $$a=\text{const.}\,  \omega_{2N-N_f}^{k}\Lambda;\quad 
k=1,\dots,2N-N_f,$$ where $\omega_{2N-N_f}$ is the $2N-N_f$-th root of unity.    If we consider two roots $a$ and $a'$ such that 
$a'=\omega_{2N-N_f}^{j}a$ for some integer $j$, we have from Eq.~(\ref{KArel})
 \be \sum_{i\geq0}\frac{\langle\Tr\Phi^i\rangle(a')}{(z')^{i+1}}=\frac{1}{\omega_{2N-N_f}^{j}}
\sum_{i\geq0}\frac{\langle\Tr\Phi^i\rangle(a)}{z^{i+1}},  \qquad  {\rm or} \quad   \sum_{i}\frac{U_i(a')}{(z')^i} =  \sum_{i}\frac{U_i(a)}{z^i}\;, 
\ee
where we have defined $z'=\omega_{2N-N_f}^{j}z$.  Eq.~(\ref{SWcurve}) tells then that the SW curve factorizes in the 
same way in all $N-r$ vacua of a given $r$, and in all vacua with $r'=N_{f}-r$. Since we know that in all $r$ vacua with 
$r<N_f/2$ the low energy physics can be described as an infrared free SQCD with r colors for any value of $m$ (see the above discussion), 
we know from \cite{APS} that the SW curve factorizes precisely as in (\ref{fac}). This proves our claim.
% the two vacua are characterized by the same low energy effective theory. We do not have a rigorous proof of this 
%fact for the massive case, but we believe that this statement is completely general.  

We have also checked the above relations, for generic $m$,  in the case of $SU(5)$  theory with $N_{f}=6$. We have verified 
by solving Eqs.~(\ref{SWcurve})-(\ref{eqxS}) with Mathematica that both $r=2$ and $r=4$ classical vacua are described by the 
identical singularity of the SW curve
\be   y^{2}=  P_N(z)^2-4\Lambda^{2N-N_f}(z+m)^{N_f} \sim   (z+m)^{4}\;,
\ee
corresponding to the quantum $r=2$ theory of Section~\ref{quantumr}. The expression for $P_{5}(x)$ from equation (\ref{SWcurve}) is rather  
lengthy and we shall  not write it. However, there are two limiting cases worth mentioning: the massless and the semiclassical 
($\Lambda\rightarrow0$) one. In the massless limit we get the following four solutions for $P_{5}(x)$:
\begin{align}
 & x^5-\sqrt{\frac{4}{3}}i\Lambda^2x^3\pm\frac{16}{9}\left(-\frac{1}{3}\right)^{1/4}\Lambda^3x^2,\nonumber\\
 & x^5+\sqrt{\frac{4}{3}}i\Lambda^2x^3\pm\frac{16}{9}\left(-\frac{1}{3}\right)^{1/4}i\Lambda^3x^2.\nonumber
\end{align}
One can easily check that the SW curve satifies the factorization condition (\ref{fac}) with $r=2$. In the $\Lambda\rightarrow0$ limit 
we expect instead to recover the semiclassical result: three $r=2$ classical vacua ($-m$ is a root of $P_{5}(x)$ with multiplicity
two) and one $r=4$ vacuum ($-m$ is a root of $P_{5}(x)$ with multiplicity four). Defining $z=x+m$ we find infact the four solutions
$$\left(z-\frac{5}{3}m\right)^3z^2,\quad\left(z-\frac{5}{3}m\right)^3z^2,\quad\left(z-\frac{5}{3}m\right)^3z^2,\quad (z-5m)z^4.$$
Notice that this result is obtained discarding all subleading terms (higher orders in $\Lambda/m$); in the exact solution all the 
polynomials are divisible just by $z^2$. The point is that the coefficients for the cubic and quadratic terms in $z$ for the fourth 
polynomial are negligible in this limit.
%\subsection{$r=2$ and $r=4$ vacua in $SU(5)$ theory with $N_{f}=6$}

%We have verified the above argument by explicitly factorizing the Seiberg-Witten curve in the case of ${\cal N}=2$,  $SU(5)$ theory with six flavors of quark hypermultiplets. 

\section{Singular points and colliding $r$ Vacua}

The physics of the local $r$ vacua represents a beautiful example of confining vacuum which is  dual Higgs system of nonAbelian variety. But perhaps even more interesting is the situation in which singular SCFT's recently studied are deformed by an ${\cal N}=1$ adjoint scalar mass term $\mu\, \Tr \Phi^{2}$. 
In this section we discuss a few examples of the vacua of this type.

\subsection{SCFT point in the massless  $USp(2N)$ theory and the  Gaiotto, Seiberg, Tachikawa (GST) dual}

It was pointed out in \cite{EHIY} that in the massless limit of ${\cal N}=2$,  $USp(2N)$ theory with $N_{f}=2n$ matter hypermultiplets new SCFTs emerge, different from those seen in $SU(N)$ theory. Note that in the massive, equal mass ($m_{i}=m \ne 0$) theory, the SCFT vacua occurring in the  $USp(2N)$ theory are the same  $r$ vacua of $SU(N)$ theory, $r=0,1,2,\ldots, N_{f}/2$, exemplifying the universality of SCFTs.   In the $m_{i}\to 0$ limit,  however, the $r$ vacua collapse into a singular SCFT  
(``Tchebyshev''  point) \cite{CKM}   with a larger global  symmetry $SO(2N_{f})$.
A recent study \cite{Simone} by one of us, done following closely the analysis by Gaiotto, Seiberg, Tachikawa \cite{GST}, has shown that the relevant SCFT can be analyzed by introducing two different scalings for the scalar VEVs $u_{i}\equiv \langle \Phi^{i} \rangle$ (the Coulomb branch coordinates)
around the singular point.     
%\be  u_{i}\sim\epsilon_{B}^{2i},  \quad ({\small i=1,\dots,N-n+2);  \qquad u_{N-n+2+i}} \sim\epsilon_{A}^{2+2i}, \quad (i=0,\dots,n-2),
%\ee
%($N_{f}=2n$)  such that  $\epsilon_{B}^{2N+4-2n}=\epsilon_{A}^2.$   
The infrared physics of this system is  \cite{Simone}  
\begin{description}
\item [(i)] $U(1)^{N-n}$ Abelian sector, with massless particles charged under each $U(1)$ subgroup.
\item [(ii)]  The  (in general, non-Lagrangian) A sector  with global symmetry $SU(2)\times SO(4n)$.
\item [(iii)] The B sector is free and describes a doublet of hypermultiplets. The flavor symmetry of this system is $SU(2)$.
%the Coulomb moduli coordinate now includes  $u_{1}$.  We interpret this as representing a low energy effective $U(1)$ gauge field.
\item [(iv)] $SU(2)$ gauge fields coupled weakly  to the last two sectors.
\end{description} 

For general $N_{f}$  these involve non-Lagrangian SCFT theories, and it is not easy to analyze the effects of $\mu\, \Tr\, \Phi^{2}$ deformation.  
 In a particular case $n=2$  ($USp(2N)$ theory with $N_{f}=4$), however,   the A sector becomes free and describes four doublets 
 of $SU(2)$.  Let us consider the effect of $\mu \, \Phi^{2}$ deformation of this particular  theory focusing on the nonAbelian
 sector. 
% The analysis of the $U(1)^{N-n}$ sector was already given in \cite{APS,CKM} and is by now standard.
The superpotential  for a hypermultiplet $Q_{0}$ and four hypermultiplets  $Q_{i}$'s, coupled to $SU(2) \times U(1)$  gauge fields  (only $Q_{0}$ carrying the $U(1)$ charge)   is
\beq\label{vac}   Q_{0} A_{D} {\tilde Q}^{0} +  Q_{0} \phi \,{\tilde Q}^{0} + \sum_{i=1}^{4}  Q_{i} \phi \, {\tilde Q}^{i}  +  \mu A_{D} \Lambda  + \mu\, \Tr \,  \phi^{2}\;.   
\eeq
The vacuum equations are 
\beq    Q_{0} {\tilde Q}^{0} + \mu \Lambda =0\;;   \label{eq1}
\eeq
\beq   (\phi + A_{D})  {\tilde Q}^{0}= Q_{0} \, (\phi + A_{D}) =0\;;  \label{eq2}
\eeq
\beq   \frac{1}{2}  \sum_{i=1}^{4}  Q_{i}^{a}   {\tilde Q}_{b}^{i} -  \frac{1}{4}   ( Q_{i}{\tilde Q}^{i} )\,  \delta_{b}^{a}  +  \frac{1}{2} Q_{0}^{a}{\tilde Q}^{0}_{b}- \frac{1}{4}  ( Q_{0}{\tilde Q}^{0} )
\, \delta^{a}_{b}  + \mu \, \phi^{a}_{b}=0\;;  \label{eq3}
\eeq
\beq    \phi \, {\tilde Q}^{i}=  Q_{i}\, \phi =0, \quad \forall i\;.   \label{eq4}
\eeq
The first tells that $Q_{0}\ne 0$.  By gauge choice
\beq  Q_{0}^{1} =  {\tilde Q}^{0 \, 1}=  \sqrt{-\mu \Lambda} \ne 0; \qquad   Q_{0}^{2} =  {\tilde Q}^{0\, 2}=0.    \label{sol1}
\eeq
A solution with   \[ \phi =  \left(\begin{array}{cc}a & 0 \\0 & -a\end{array}\right)  \ne 0\] would necessarily imply
\beq    Q_{i}= {\tilde Q}^{i} =0\;; \quad \forall i,\qquad A_{D}=-a\;,  \quad  a= \frac{\Lambda}{4}\;, 
\eeq
Such a solution involves a fluctuation ($\sim\Lambda$)  beyond the validity of the effective action:  it is an artefact of the low-energy effective action and must be disregarded. 
We must therefore choose
\beq   \phi =0, \quad  A_{D}=0\;.\label{sol2}
\eeq 
The contribution from $Q_{i}$'s must then cancel that of   $Q_{0}$ in  Eq.~(\ref{eq3}).   
By flavor rotation  the nonzero VEV can be attributed to $Q_{1}, {\tilde Q}^{1}$, i.e., either of the form
\beq     (Q_{1})^{1} =  ({\tilde Q}^{1})_{1} = \sqrt{\mu \Lambda} \;,    \qquad  Q_{i}={\tilde Q}_{i}=0, \quad i=2,3,4.    \label{sol3a}
\eeq
or  
\beq     (Q_{1})^{2} =  ({\tilde Q}^{1})_{2}   = -  \sqrt{\mu \Lambda} \;,    \qquad  Q_{i}={\tilde Q}_{i}=0, \quad i=2,3,4. \label{sol3b}
\eeq
The  vacuum  is thus given by Eq.~(\ref{sol1}), Eq.~(\ref{sol2})  and   Eq.~(\ref{sol3a}) or Eq.~(\ref{sol3b}).
This means that the flavor symmetry is broken as
\beq   SO(8) \to    U(1) \times SO(6) = U(1) \times SU(4), 
\eeq
where the $U(1)$ factor is the color-flavor diagonal $SO(2)$ left unbroken by Eq.~(\ref{sol3a}) or Eq.~(\ref{sol3b}).  
This pattern of symmetry breaking is precisely what is  expected from the result known at large $\mu \gg \Lambda$ \cite{CKM}, showing the  consistency of the whole picture.

\subsection{Singular points in $SU(N)$ SQCD}

In this section we study two types of singular points in $SU(N)$ SQCD (with an even number   $N_f=2n$ of  flavors) which are relevant for the breaking to $\mathcal{N}=1$. 
In the first case the singular points arise   from the collision of different $r$ vacua. Dynamical flavor symmetry breaking does 
not occur in this case. The second class of singular points arise in a ``degeneration limit'' of $r=n$ vacua, in 
which the SW curve becomes more singular. Vacua with different $r$ are not involved in this case and the pattern of flavor symmetry 
breaking remains to be $U(N_f)\rightarrow U(n)\times U(n)$.

\subsubsection{Colliding $r$-vacua of  the  $SU(N)$ SQCD  \label{collision}  }

It has been noted recently \cite{noi,Simone} that when the (equal) quark mass parameter is fine-tuned to a particular value of the order of $\Lambda$,
\be   m=m^{*}\equiv   \omega^{k}\, \frac{2 N - N_{f}}{N}\, \Lambda\;, \qquad (k=1,\ldots,  2N-N_{f}, \quad  \omega^{2N- N_{f}}=1)\;,   \label{collidingvacua}
\ee
 all the $r$-vacua with $r=0,1,\ldots, \tfrac{N_{f}}{2}$ (more precisely, one representative from each $r$ vacua) coalesce to form a single vacuum \footnote{In \cite{Bolo}
 an analogous phenomenon was studied, but by using an appropriate ${\cal N}=1$ superpotential $W(\Phi)$ and selecting particular vacua.}
 \be     y^{2}\sim   (x+m^{*})^{N_{f}+1}\;.
\ee
This corresponds to the SCFT of the highest criticality  \cite{EHIY}  (called EHIY point in \cite{GST})  for  
\be   N= n+1 \; \qquad (N_{F}=2n) \;.  \label{collapse}
\ee   
 Now the  EHIY points  in general $SU(N)$  theories with  $N_{F}=2n$   flavors have recently been  reanalyzed  by  Gaiotto, Seiberg and Tachikawa \cite{GST}.  
 According to these authors, the low-energy system is an infrared free $SU(2)$ gauge theory, weakly coupled to two separate SCFT's. One (the A sector)  is a strongly coupled (in general, without local Lagrangian description) theory with $SU(2) \times SU(N_{f})$ flavor symmetry,  
 the other  (the sector B)  is the most singular superconformal point of $SU(N-n+1)$ theory with two flavors, 
with $SU(2)\times U(1)$ flavor symmetry.  The diagonal combination of the $SU(2)$ flavors is weakly gauged.  
It is easy to see that the low energy physics at the singular point of interest for us can be described as \cite{Simone}
\begin{description}
\item [(i)] A $U(1)^{N-n-1}$ Abelian sector, with massless particles charged under each $U(1)$ subgroup.
\item [(ii)]  The  (in general, non-Lagrangian) A sector  with global symmetry $SU(2)\times SU(N_f)$.
\item [(iii)] The B sector is the most singular superconformal point of $SU(2)$ theory with two flavors (or the $D_3$ Argyres-Douglas
theory), with $SU(2)\times U(1)$ flavor symmetry.
%the Coulomb moduli coordinate now includes  $u_{1}$.  We interpret this as representing a low energy effective $U(1)$ gauge field.
\item [(iv)] $SU(2)$ gauge fields coupled to the last two sectors.
\end{description} 
%As  these $SU(2)$ gauge interactions are infrared free,  we would have 
%\be      \langle  \lambda \lambda  \rangle  =0\;, 
%\ee
%without the $\mu \Phi^{2}$ perturbation. 
The presence of the $\mu \Phi^{2}$ term, breaking $SU_{R}(2)$ explicitly, is expected to generate nonvanishing gaugino condensate through anomaly, and induce the symmetry breaking 
\be   {\mathbbm Z}_{2N-N_{f}} \to   {\mathbbm Z}_{2}\;. 
\ee
We are not able to deduce such a result directly  with the  GST dual description. 
%let us return to the special case of the collapsed $r$-vacua, with (\ref{collapse}), that is, an $SU(N)$ theory with $N_{f}=2N-2$.   In this case,  the B sector is the most singular SCFT of  $SU(N-n+1)=SU(2)$  theory with two flavors with global symmetry $SU(2)\times U(1)$, studied in \cite{EHIY}.    Furthermore, if we specialize in the case of 
%$N=3$, $N_{f}=4$ theory,  
%the A sector becomes free and describes three doublets of $SU(2)$ with global symmetry, $SO(6)\sim SU(4)$.  
However, what happens in the colliding $r$ vacua in general $SU(N)$ gauge theories with even $N_{f}$ flavors,
  perturbed by  the adjoint mass term, can be exactly  determined by the generalized Konishi anomaly relations
\cite{CDSW} and  by use of the Dijkgraaf-Vafa superpotential.   The analysis of \cite{noi} shows that the meson  and gaugino condensates are of the form (see Eq.~(\ref{eqxa}) and Eq.~(\ref{eqxS}))
\be    \langle\tilde{Q}^{i}Q_{i}\rangle \sim \, \mu \Lambda\,, \quad ({\rm indep. ~ of} \,\,  i)\;;  \qquad  \langle  \lambda \lambda  \rangle  \sim \mu \Lambda \ne 0\;. 
\ee
The $SU(N_{f})\times U(1)$ symmetry remains unbroken, in contrast to what happens (for generic $m$)  in single $r$ vacua, (\ref{symbr}). 
%It would be an interesting problem to understand how the nonLagrangian theories in the GST  dual description 
%give rise to these condensates upon the ${\cal N}=1$ deformation. 

\subsubsection{Higher order singularity  \label{non collision}}

The vacuum arising from the collision of r vacua is not the only higher singular point in softly broken $\mathcal{N}=2$ $SU(N)$ SQCD.
For illustration let us consider the simplest example, namely $SU(4)$ theory with $N_f=4$: the SW curve is
$$y^{2}=(x^4-u_2x^2-u_3x-u_4)^{2}-4\Lambda^{4}(x+m)^4.$$ In this case the $r=2$ vacuum can be found easily and the curve assumes the form
$$y^2=(x+m)^{4}(x-m)^{2}(x-m-2\Lambda)(x-m+2\Lambda).$$ From here we easily see that when $m=\pm\Lambda$ the curve can be approximated as
$y^2\approx(x+m)^5$ and we recover the case studied before. On the other hand, in the limit $m=0$ the curve becomes more singular and 
reduces to $y^2\approx x^6$. Of course, this singular point exists in the general case, as long as $n<N-1$ (this was already noticed in \cite{CKM}):
in an  $r=\frac{N_f}{2}$ vacuum the SW curve assumes the form $$y^2=(x+m)^{N_f}Q_{N-n-1}^{2}(x)(x-\alpha)(x-\beta).$$ The roots 
of $Q_{N-n-1}$ have multiplicity one and are located at (see \cite{CKM}) 
\be   x=\frac{N_f}{2N-N_f}m+2\Lambda\cos\left(\frac{2k\pi}{2N-N_f}\right);\quad k=1,\dots,N-\frac{N_f}{2}-1.   \label{tcheb}\ee
When the bare mass is chosen in such a way that $-m$ coincides with one of these roots the SW curve can be approximated as $y^2\approx(x+m)^{N_f+2}$.

From the analysis performed in \cite{GST} we can conclude that the low energy physics at this singular point can be described as follows:
\begin{description}
\item [(i)] A $U(1)^{N-n-2}$ Abelian sector, with massless particles charged under each $U(1)$ subgroup.
\item [(ii)]  The A sector  with global symmetry $SU(2)\times SU(N_f)$ described in the previous section.
\item [(iii)] The B sector is the most singular point of $SU(3)$ theory with two flavors (or the $D_4$ Argyres-Douglas
theory), with $SU(2)\times U(1)$ flavor symmetry \footnote{Actually, it was recently shown in \cite{AMT} that in this case the 
flavor symmetry enhances to $SU(3)$. However, an $SU(2)$ subgroup is gauged and the manifest flavor symmetry is the commutant 
of $SU(2)$ inside $SU(3)$, which is $U(1)$.}.
%the Coulomb moduli coordinate now includes  $u_{1}$.  We interpret this as representing a low energy effective $U(1)$ gauge field.
\item [(iv)] $SU(2)$ gauge fields coupled to the last two sectors.
\end{description} 

From the analysis performed in \cite{noi} it is easy to see that the pattern of symmetry breaking (once the superpotential for the adjoint field is
 turned on) is $U(N_f)\rightarrow U(N_f/2)\times 
U(N_f/2)$, the same as in the $r=N_f/2$ vacuum 
 In contrast to the case discussed in the previous paragraph  \ref{collision}, this vacuum does 
not arise from the coalescence of different r vacua.

\subsubsection{Breaking to $\mathcal{N}=1$ in the singular vacua}

In this section we wish  to test our proposal for the low-energy effective description at the singular points by reproducing the 
correct pattern of flavor symmetry breaking occurring once the $\mu\Tr\Phi^2$ perturbation is turned on.

As we have seen, for generic $m$ the most singular point in the moduli space is the $r=N_f/2$ vacuum \footnote{As is well known, 
the curve can become even more singular. However, such points are not relevant from the point of view of the breaking to $\mathcal{N}=1$.} (in 
which the SW curve can be approximated as $y^2\approx(x+m)^{N_f}$). Its low-energy physics 
involves a scale invariant sector with $SU(N_f/2)$ gauge group, whose coupling constant depends on $m$. For 
special values $m^{*}$ of order $\Lambda$ (or zero) the SW curve degenerates further ($y^2\approx(x+m)^{N_f+1}$ or $y^2\approx(x+m)^{N_f+2}$,
as we have seen above). It is easy to see that, as we approach these critical values the coupling constant of the $SU(N_f/2)$ 
theory diverges \cite{ShapTach}. In this limit it is convenient to adopt the Argyres-Seiberg dual description \cite{AS} in which an
 $SU(2)$ gauge group emerges, coupled to a hypermultiplet in the doublet (B sector) and to a strongly coupled interacting sector 
 which coincides precisely with the A sector introduced above.
 
In order to describe the low energy physics at the most singular point in a neighbourhood of the critical values $m^{*}$, it is 
thus convenient to introduce these two sectors. As we approach the critical value $m^{*}$ the curve becomes more singular and the B sector, 
which is free in the $r$ vacuum, becomes interacting ($D_3$ or $D_4$ Argyres-Douglas theory for the two classes of 
singular points we are interested in). In the process the A sector is just a spectator.

%for particular values $m^*$ of the bare mass there are higher singular points whose low energy physics involves two sectors. This
%kind of description is valid also in the $r=N_f/2$ vacuum for values of $m$ close to the critical ones ($m-m^*\ll\Lambda$): 
%at the level of the SW curve it is easy to see that this perturbation affects only the B sector, which becomes free and describes a doublet of 
%$SU(2)$. The A sector is just a spectator. This is not surprising, since the low energy physics in the r vacuum is that of $U(n)$ theory with $2n$ 
%flavors, whose coupling constant diverges in the singular limit $m=m^*$. In this regime we can adopt the Argyres-Seiberg dual description, which involves 
%precisely the A sector coupled to a doublet of $SU(2)$. This is precisely what we get from the curve.

Finding the effective low energy description at these singular points once the $\mu\Tr\Phi^2$ term is turned on is in general rather difficult. However, 
in the $N_f=4$ case the problem is greatly simplified, since the A sector is free. For $m$ close to $m^*$ the low energy physics in the 
$r=2$ vacuum admits a Lagrangian description analogous to (\ref{vac}). The only difference is that the A sector describes three 
doublets of $SU(2)$ instead of four. Imposing the F-term equations we find as before a non vanishing condensate for the $Q_{0}$ and $Q_1$ fields
 (we use the same notation as in (\ref{vac})),
reproducing the correct pattern of symmetry breaking: $$U(1)\times SO(6)\rightarrow U(1)\times U(1)\times SO(4)\simeq U(2)\times U(2).$$
Clearly, if the condensate for $Q_0$ vanishes, the one for $Q_1$ vanishes as well, restoring the full $U(N_f)$ flavor symmetry of 
the theory. The $Q_0$ condensate can be determined focusing on the B sector only, which is the most singular point in $SU(2)$ or 
$SU(3)$ theory with two flavors in the cases of interest for us. The problem is thus reduced to computing the Abelian condensates 
in $SU(2)$ or $SU(3)$ theories.
The result of the direct analysis (see Appendix \ref{23}) is that the $Q_0$ condensate vanishes in the $SU(2)$ case but not in the $SU(3)$ one, reproducing the expected pattern of flavor symmetry breaking discussed in the two paragraphs \ref{collision} and \ref{non collision}, respectively.

\section{Discussion} 

Altogether,  what we learned from the softly broken ${\cal N}=2$ SQCD, with quark supermultiplets in the fundamental representation, supports a fairly standard picture of confinement: magnetically charged particles condense, leading to quark confinement \cite{NM}. The details of how this occurs, however,  depends on the parameters and particular vacuum considered in various  ways, and as soon as we stray away from the well understood cases of Abelian-dual-superconductor systems, we appear suddenly to find ourselves in a foreign land full of unfamiliar phenomena.  

The physics of $r$ vacua for small masses $m$ in the $SU(N)$ SQCD  is characterized by the (dual-) color-flavor locking condensates of nonAbelian magnetic monopoles. The dual gauge symmetry is completely broken (confinement), at the same time leaving the global $SU(r)$ symmetry intact. 
This echos the properties of the color-flavor locked vacuum found at the $r$ (or $N_{f}-r$) semiclassical singularity, for large $m, \mu$, with the same global symmetry. 
Such a matching is absolutely indispensable to realize a complete Higgsing of the dual gauge symmetry -  confinement - yet keeping the same global symmetry as the corresponding semiclassical theory in Higgs phase. 
The flow from the semiclassical (large $m \gg \mu \gg \Lambda$) to  quantum  ($m \sim \mu \ll \Lambda$) regions is smooth as  our theory involves scalars in the fundamental representation (another way of seeing it is the fact that the ${\cal N}=1$ supersymmetry maintained throughout guarantees a holomorphic dependence on 
$m$ and $\mu$ so that no discontinuous change of physics is possible between the two regions). 

The ``nonAbelian vortex'' carrying the $CP^{r-1}$ moduli fluctuations in the quantum $r$ vacuum, Eq.~(\ref{eq:sigmamodelaction}),  is nothing but the fluctuations of the chromoelectric,  confining string flux:  the gauge symmetry is dynamically broken to $SU(r) \times U(1)^{N-r}$. The system thus presents a rare instance in which the properties of the confining string can be analytically studied, thanks to the existence of a weakly coupled dual (but still local) description.

A more intriguing situation occurs when the bare mass is tuned to a special value of the order of $\Lambda$,   $m \to m^{*}$.  For the particular choice of the critical mass, Eq.~(\ref{collidingvacua}),  all the $r$ vacua   ($r=0,1,\ldots, N_{f}/2$)  collide to form a single vacuum, which corresponds to one of the singular SCFT  which have been given much attention recently, starting from the work by Argyres and Seiberg.  
The change of the global symmetry in the limit of coalescing vacua, from $U(r) \times U(N_{f}-r)$ to  $U(N_{f})$,  indicates a different set of condensates inducing confinement,  i.e., a new mechanism of confinement.   The system is now described as a deformation of a   
singular, non-Langrangian  SCFT;   the direct analysis of this phenomenon in terms of the low-energy degrees of freedom, is still beyond our ability,  even though the  
exact answer on the symmetry breaking pattern can be inferred from the ${\cal N}=1$ side, via the generalized Konishi anomaly and the Dijkgraaf-Vafa superpotential \cite{noi}.  

For a still another choice of the critical mass, Eq.~(\ref{tcheb}),  the highest  $r$ vacua, with $r=\tfrac{N_{f}}{2}$ becomes more singular, without however colliding with other lower $r$ vacua.  In this case the pattern of symmetry breaking remains to be  $U(\tfrac{N_{f}}{2}) \times U(\tfrac{N_{f}}{2}).$ 

We have given,  for particular cases of $N_{f}=4$ theories, justifications for these results \cite{CKM,Simone}  by using the direct analysis of the singular vacua  \cite{AS, GST}.   

Finally,  in $USp(2N)$ theory the collision of the $r$ vacua occurs when the bare quark mass approaches zero, when the global symmetry of  
the underlying theory gets enhanced from  $SU(N_{f})\times U(1)$ to $SO(2N_{f})$.  A study of the corresponding semi-classical system (at large  $\mu \gg \Lambda $) shows that    
the global symmetry is broken to $U(N_{F})$, indicating a new set of condensates forming in the limit.  In the case of $USp(2N)$ theory with $N_{f}=4$, where the GST dual becomes sets of free hypermultiplests weakly coupled to $SU(2)$ gauge fields,  we have been able to reproduce such a result directly from the GST dual description, showing the consistency of our reasonings. 

To conclude,  a proper description of the conformal theories appearing as the infrared fixed-point theories
as found recently by Seiberg et. al. \cite{AS,GST}, is indispensable for the understanding of confinement,  because   in systems discussed here
the latter occurs as the result of small deformation of conformally invariant systems.
This means that the nature of the degrees of freedom and their interactions in the conformal limit determine how confinement is induced by the deformation.  

A series of papers by Shifman and Yung have discussed many related issues  in $U(N)$ theories \cite{SYrec}.

\appendix

\section{Classical vacuum equations}

The superpotential has the form
\beq W= \mu \Tr \, \Phi^2  + \sqrt2 \,{\tilde Q}_i^a \Phi_a^b Q_b^i +
     m_i \, {\tilde Q}_i^a Q_a^i\;.\eeq
The vacuum equations read
\beq   [ \Phi, \Phi^{\dagger}] = 0 \, ;
\label{D1}
\eeq
\beq
\nu \delta_a^b=  Q_a^i (Q^{\dagger})_i^b - ({\tilde Q}^{\dagger})_a^i
{\tilde Q}_i^b \, ;
\label{D2}
\eeq
\beq
Q_a^i {\tilde Q}_i^b - \frac{ 1}{ N} \delta_a^b (Q_c^i {\tilde Q}_i^c) +
\sqrt2  \, \mu \Phi_a^b = 0 \, ;
\label{F1}
\eeq
\beq
Q_a^i m_i + \sqrt2 \,\Phi_a^b Q_b^i = 0  \qquad ( { \hbox {\rm no sum
over}} \, \,i) \, ;
\label{F2}
\eeq
\beq
m_i {\tilde Q}_i^a +  \sqrt2 \, {\tilde Q}_i^b \Phi_a^b = 0 \qquad (
{\hbox{\rm no sum over}} \,\,i) \, .
\label{F3}
\eeq
By gauge rotation $\Phi$ can be taken as
\beq
\Phi = \diag \, (\phi_1, \phi_2, \ldots  \phi_{N}) \, , \qquad \sum
\phi_a = 0 \, .
\label{phivev}
\eeq
  $Q_a^i $ and ${\tilde  Q}_i^b$
are either nontrivial eigenvectors of the matrix $\Phi $ with possible
eigenvalues $m_i$,  or null vectors.    The solution with eigenvalues $m_1, m_2, \ldots, m_r$ is    
\beq
\Phi = \frac{1}{\sqrt{2}}\diag \, (-m_1, -m_2, \ldots, -m_r, c,
\ldots , c) \, ; \qquad c = \frac{1 }{ N-r} \sum_{k=1}^r m_k \, .
\label{diagphi}
\eeq
\beq   Q_a^i  =
\left(\begin{array}{cccc}   f_1 &  &  &  \\    & \ddots &    &  \\  &  & f_r &      \\   &  &    & {\mathbf 0}\end{array}\right)\;, \qquad {\tilde Q}_i^a  =
\left(\begin{array}{cccc}{\tilde f}_1 &  &  &  \\    & \ddots &    &  \\  &  & {\tilde f}_r &      \\   &  &    & {\mathbf 0}\end{array}\right)\;, 
\label{vevofqti}\eeq
where
\beq   r=0,1, \ldots, {\hbox {\rm min}} \, \{N_f, N-1\},
\eeq
The solution for $f_{i}, {\tilde f}_{i}$ is   (see \cite{CKM} for more details)
\beq
f_i {\tilde f}_i =  \, \mu \,m_i  + \frac{ 1 }{ N -r} \mu \,\sum_{k=1}^r
m_k \;, \qquad  
f_i^2 = | {\tilde f}_i |^2 \,, \qquad 
(f_{i} > 0)\, .
\label{solnd}
\eeq
The number of the quark flavors ``used'' to make solutions define  various classical $r$-vacua. 
As the  solution with a given $r$ leaves a local  $SU(N-r)$ invariance  it counts as a set of $N-r$  solutions (Witten's
index).
In all  there are precisely
\beq
{\cal N} = \sum_{r=0}^{{\hbox {\rm min}} \, \{N_f, N-1\}}\, (N-r)
\, \binom{N_f  }{  r}
\label{nofvac}
\eeq
classical  solutions for generic $m_{i}$'s and $\mu \ne 0$.
%(For $r=0$, $Q_a^i = {\tilde Q}_i^b =0,$ $\Phi=0$ is obviously a
%solution with full $SU(N)$ invariance.)

\section{Equations determining VEVs in the quantum  $r$ vacua}

The D-tem potential gives 
\beq  0=  [ \phi, \phi^{\dagger}]; \label{effD1}\eeq
\beq \nu \delta_a^b=  q_a^i (q^{\dagger})_i^b - ({\tilde
q}^{\dagger})_a^i
   {\tilde q}_i^b; \label{effD2}\eeq
\beq 0=  q_a^i (q^{\dagger})_i^a - ({\tilde
q}^{\dagger})_a^i
   {\tilde q}_i^a; \label{effD2bis}\eeq
   while the F-term equations are 
\beq q_a^i {\tilde q}_i^b - \frac{ 1}{r} \delta_a^b (q_c^i {\tilde q}_i^c)
+ \sqrt2  \, \mu \phi_a^b =0; \label{effF1} \eeq
\beq   0= \sqrt2 \,\phi_a^b q_b^i + q_a^i  (m_i  + \sqrt2  \, a_{D0})
; \quad ( {\hbox{\rm no
sum
over}} \,\,i,\,\, a) \,
   \label{effF2} \eeq
\beq  0=  \sqrt2 \, {\tilde q}_i^b
\phi_b^a +   (m_i  + \sqrt2  \, a_{D 0}) \,  {\tilde q}_i^b
\quad ( {\hbox{\rm no sum over}}\,\,i, \,\, a).  \, \label{effF3} \eeq
\beq  \sqrt2 \, \Tr ( q {\tilde q})  + \mu \Lambda =0.  \label{effF4} \eeq
The $SU(r)$  adjoint  scalars can be diagonalized  by color rotations,
\beq   \diag \, \phi = ( \phi_1, \phi_2, \ldots  \phi_{r}), \qquad
\sum \phi_a=0. \label{effphivev} \eeq

\section{Non-Abelian vortex in the $r$ vacua}

\begin{align}
{\cal M} &=
\begin{pmatrix}
e^{i\theta}\phi_1(\rho)    1 & 0 \\
0 & \phi_2(\rho)\mathbf{1}_{r-1}
\end{pmatrix}
= \frac{e^{i\theta}\phi_1(\rho)+\phi_2(\rho)}{2}\mathbf{1}_{r}
+ \frac{e^{i\theta}\phi_1(\rho)-\phi_2(\rho)}{2} T \;,   \nonumber   \\
A_i &=
\frac{1}{2}\epsilon_{ij}\frac{x^j}{\rho^2}\left[
\left(1-f(\rho)\right) \mathbf{1}_{r}+\left(1-f_{\rm NA}(\rho)\right) T
\right]  \ ,     \label{minivort1}
\end{align}
which is oriented to a specific direction.  In (\ref{minivort1})
\beq T \equiv \diag\left(1,-\mathbf{1}_{r-1}\right)\ , \eeq
and $z, \rho,\theta $ are cylindrical coordinates.  
The profile functions $\phi_{1,2}(\rho), \, f(\rho), f_{\rm NA}(\rho)$ satisfy the boundary conditions 
\begin{align}
\phi_{1,2}(\infty) = \frac{v}{\sqrt{2N}} \ , \quad
f(\infty)=f_{\rm NA}(\infty) = 0 \ , \quad
\phi_1(0) = 0 \ , \quad
\p_r\phi_2(0) = 0 \ , \quad
f(0) = f_{\rm NA}(0) = 1 \ .
\label{eq:bcs}
\end{align}
The vortex oriented in a generic direction in color-flavor space can be written as 
\begin{align}
{\cal M} &= U
\begin{pmatrix}
\phi_1(\rho)  1 & 0 \\
0 & \phi_2(\rho)\mathbf{1}_{r-1}
\end{pmatrix} U^{-1}
= \frac{\phi_1(\rho)+\phi_2(\rho)}{2}\,  \mathbf{1}_{r}
+ \frac{\phi_1(\rho \rho)-\phi_2(\rho)}{2} \,U T U^{-1} \ , \non
A_i &= -\frac{1}{2}\epsilon_{ij}\frac{x^j}{r^2} \left[
f(\rho)\, \mathbf{1}_{r} + f_{\rm NA}(\rho) \, U T U^{-1} \right] \ ,
\qquad  i=1,2 \ . \label{genericorient}
\end{align}
The matrix $U$ represents the coset  
\be     SU(r) / SU(r-1)\times U(1) \sim CP^{r-1},
\ee 
and is expressed in terms of an $r-1$ dimensional complex vector $B$ as 
\begin{align}
U =
\begin{pmatrix}
1 & - B^\dag \\
0 & \mathbf{1}_{r-1}
\end{pmatrix}
\begin{pmatrix}
X^{-\frac{1}{2}} & 0 \\
0 & Y^{-\frac{1}{2}}
\end{pmatrix}
\begin{pmatrix}
1& 0 \\
B & \mathbf{1}_{r-1}  
\end{pmatrix}
=
\begin{pmatrix}
X^{-\frac{1}{2}} & - B^\dag Y^{-\frac{1}{2}} \\
B X^{-\frac{1}{2}} & Y^{-\frac{1}{2}}
\end{pmatrix}
\label{eq:Umatrix} \ ,
\end{align}
where the matrices $X$ and $Y$ are defined by
\beq
X\equiv   1+ B^\dag B \ , \quad
Y\equiv\mathbf{1}_{r-1} + B B^\dag \ ,
\eeq
This form of the unitary $SU(r)$ matrices containing only the  coset coordinates $B$ is known as the reducing matrix. 

\section{Monopole condensates in $SU(2)$ and $SU(3)$ $N_f=2$ theories}\label{23}

The SW curve for the $SU(2)$ theory with two flavors can be written as $$y^2=(x^2-u)^2-4\Lambda^2(x+m)^2,$$ and if we set $u=m^2$ it 
degenerates to \beq\label{22f} y^2=(x+m)^2(x-m-2\Lambda)(x-m+2\Lambda).\eeq The low energy physics at this point is described by an Abelian $U(1)$ theory with 
two massless electrons and when we turn on the $\mathcal{N}=1$ deformation $\mu\Tr\Phi^2$, the corresponding effective action includes the superpotential 
$$\sqrt{2}\tilde{Q}_1A Q_1+\sqrt{2}\tilde{Q}_2A Q_2+\mu U; \quad U\equiv\langle\Tr\Phi^2\rangle.$$ The equations of motion thus impose the constraint 
$$\langle\tilde{Q}_1Q_1\rangle+\langle\tilde{Q}_2Q_2\rangle=-\frac{\mu}{\sqrt{2}}\frac{\partial U}{\partial A}.$$ In order to compute the 
condensate we now have to evaluate $\partial U/\partial A$. This can be done noticing that $$\frac{\partial U}{\partial A}^{-1}= 
\frac{\partial A}{\partial U}=\int_{\gamma}\frac{\partial\lambda}{\partial U},$$ where the contour $\gamma$ is a small circle surrounding the point 
$x=-m$. We can now exploit the fact that the SW differential for $SU(N)$ SQCD satisfies the relation \cite{SUN} $$\frac{\partial\lambda}{\partial U}=\frac{dx}{y}
x^{N-2}.$$ From (\ref{22f}) we then obtain $$\frac{\partial U}{\partial A}\propto\sqrt{(\Lambda+m)(\Lambda-m)}.$$ This quantity vanishes for $m=\pm\Lambda$,
which are precisely the values such that the SW curve degenerates further and we encounter the $D_3$ Argyres-Douglas point. This shows that the $Q_0$ 
condensate (in the notation of \ref{vac}) vanishes at this point. 

The computation for $SU(3)$ is similar: the SW curve in this case is $$y^2=(x^3-Ux-V)^2-4\Lambda^4(x+m)^2$$ and setting 
$U=2\Lambda^2+\frac{3}{4}m^2$, $V=2m\Lambda^2-\frac{m^3}{4}$ we reach the $r=1$ vacuum, the point we are looking for. 
The SW curve at this point factorizes as \beq\label{32f}y^2=(x+m)^2(x-\frac{m}{2})^2(x^2-mx+\frac{m^2}{4}-4\Lambda^2),\eeq
and the low energy effective action describes an Abelian $U(1)^2$ theory with two massless hypermultiplets charged under 
one $U(1)$ factor and another hypermultiplet charged under the second one. In the $m\rightarrow0$ limit the curve degenerates 
further and we find the maximally singular point. In order to find the $Q_0$ condensate we have to evaluate as before
$$\frac{\partial A}{\partial U}=\int_{\gamma}\frac{\partial\lambda}{\partial U},$$ where the contour $\gamma$ is again a circle around the point 
$x=-m$. The crucial difference with respect to the $SU(2)$ case is the fact that now $$\frac{\partial\lambda}{\partial U}=\frac{xdx}{y},$$
leading to the relation $$\frac{\partial A}{\partial U}\propto\left(\sqrt{4\Lambda^2-\frac{9}{4}m^2}\right)^{-1}.$$ This quantity 
remains finite in the $m\rightarrow0$ limit (notice that $\partial A/\partial V$ diverges instead). 
The computation of $\partial U/\partial A$, the quantity we are interested in, is slightly more delicate with respect to the $SU(2)$ case, since the 
Coulomb branch has now complex dimension two and the $A$ cycle is a function of both $U$ and $V$. It is convenient to introduce the homology cycle $B$, which 
satisfies the equation
$$\frac{\partial B}{\partial U}=\int_{\gamma'}\frac{\partial\lambda}{\partial U},\quad \frac{\partial B}{\partial V}=\int_{\gamma'}\frac{\partial\lambda}{\partial V},$$ 
where $\gamma'$  is a loop around the point $x=\frac{m}{2}$. We can take $A$ and $B$ as a basis of ``electric'' cycles. From the above formulas it is 
clear that $\partial A/\partial U$ and $\partial B/\partial U$ are both finite in the massless limit, whereas $\partial A/\partial V$ and $\partial B/\partial V$
are both proportional to $\sim\frac{1}{m}$ for small $m$. Considering now the equations
$$\frac{\partial A}{\partial U}\frac{\partial U}{\partial A}+\frac{\partial A}{\partial V}\frac{\partial V}{\partial A}=1;\quad 
\frac{\partial B}{\partial U}\frac{\partial U}{\partial A}+\frac{\partial B}{\partial V}\frac{\partial V}{\partial A}=0,$$
we can easily see that they cannot be satisfied if $\partial U/\partial A$ vanishes. This guarantees that the $Q_0$ condensate does not vanish.


\begin{thebibliography}{55}

\bibitem{AS}
P. C. Argyres and N. Seiberg,
%``S-duality in $\mathcal{N}=2$ supersymmetric gauge theories,''
 JHEP\ 0712:088 (2007),
  arXiv:0711.0054 [hep-th].
  
\bibitem{GST} 
  D.~Gaiotto, N.~Seiberg and Y.~Tachikawa,
 % ``Comments on scaling limits of 4d N=2 theories,''
  JHEP {\bf 1101}, 078 (2011), 
  arXiv:1011.4568 [hep-th].
  %%CITATION = ARXIV:1011.4568;%%
 
\bibitem{G}
D. Gaiotto,
%``$\mathcal{N}=2$ dualities,''
arXiv:0904.2715 [hep-th].

\bibitem{CD}
O. Chacaltana and J. Distler,
%``Tinkertoys for Gaiotto duality,''
JHEP\ 1011:099 (2010),
  arXiv:1008.5203 [hep-th].

\bibitem{CV1}
M. Alim, S. Cecotti, C. Cordova, S. Espahbodi, A. Rastogi and C. Vafa,
%``$\mathcal{N}=2$ Quantum Field Theories and Their BPS Quivers,''
 arXiv:1112.3984 [hep-th].

\bibitem{GMN}
D. Gaiotto, G. W. Moore and A. Neitzke,
%``Spectral networks,''
 arXiv:1204.4824 [hep-th]. 
  
  \bibitem{HT}
A.~Hanany and D.~Tong,
%``Vortices, instantons and branes,''
JHEP {\bf 0307}, 037 (2003),   hep-th/0306150.

\bibitem{ABEKY}
R.~Auzzi, S.~Bolognesi, J.~Evslin, K.~Konishi and A.~Yung,
%``Nonabelian superconductors: Vortices and confinement in N = 2 SQCD,''
Nucl.\ Phys.\ B {\bf 673} (2003) 187,
hep-th/0307287.
%%CITATION = HEP-TH 0307287;%%
 
 \bibitem{ABEK}
  R.~Auzzi, S.~Bolognesi, J.~Evslin and K.~Konishi,
 % ``Nonabelian monopoles and the vortices that confine them,''
  Nucl.\ Phys.\  B {\bf 686} (2004) 119, 
  hep-th/0312233;  
 M.A.C.~Kneipp, 
  %``Color superconductivity, Z(N) flux tubes and monopole confinement in deformed N=2* superYang-Mills theories'', 
Phys. Rev. D {\bf 69}: 045007 (2004), 
hep-th/0308086.
 
 
   \bibitem{Duality}
 M.~Eto, L.~Ferretti,  K.~Konishi, G.~Marmorini, M.~Nitta, K.~Ohashi, W.~Vinci and N.~Yokoi,
  %``Non-abelian duality  from  vortex moduli:  a dual model of  color-confinement'',   
Nucl. Phys. B  {\bf 780} 161-187, 2007,
  hep-th/0611313.
   
     \bibitem{SY}
% M. Shifman and A. Yung, Phys. Rev. D {\bf 66},  045012  (2002);    
  M.~Shifman and A.~Yung,  Phys. Rev.  D {\bf 70},  045004  (2004),  hep-th/0403149;  
A.~Gorsky, M.~Shifman and A.~Yung,  Phys. Rev. D  {\bf 71}, 045010 (2005),  hep-th/0412082.
  
\bibitem{ModMat}
 M.~Eto, Y.~Isozumi, M.~Nitta, K.~Ohashi and N.~Sakai,
 %``Solitons in the Higgs phase: The moduli matrix approach,''
 J.\ Phys.\ A  {\bf 39}, R315 (2006),  hep-th/0602170. 

\bibitem{Tong}
D.~Tong, ``TASI lectures on solitons: Instantons, monopoles, vortices and kinks'', 
arXiv:hep-th/0509216,   
  ``Quantum Vortex Strings: A Review,''
 arXiv:0809.5060 [hep-th].

\bibitem{SYBook}  M.~Shifman and A.~Yung,
   Rev.\ Mod.\ Phys.\  {\bf 79}, 1139 (2007), 
 hep-th/0703267.

\bibitem{GJK} S.B.~Gudnason, Y.~Jiang and  K.~Konishi,   JHEP {\bf 1008}:012 (2010) , arXiv:1007.2116 [hep-th].


\bibitem{CKM}
 G.~Carlino, K.~Konishi and H.~Murayama,
%  ``Dynamical symmetry breaking in supersymmetric SU(n(c)) and USp(2n(c)) gauge theories,''
  Nucl.\ Phys.\  {\bf B590} (2000) 37, hep-th/0005076. 
  


 \bibitem{SW1}
N. Seiberg and E. Witten, Nucl.Phys. {\bf B426} (1994) 19; Erratum
\textit{ibid.} \textbf{B430} (1994) 485, hep-th/9407087.

\bibitem{SW2}
N. Seiberg and E. Witten, Nucl. Phys. {\bf B431} (1994) 484,
   hep-th/9408099.

\bibitem{SUN}
P.~C.~Argyres and A.~F.~Faraggi, Phys. Rev. Lett {\bf 74} (1995)
3931, hep-th/9411047;
A. Klemm, W. Lerche, S. Theisen and S. Yankielowicz, Phys. Lett.
{\bf B344} (1995) 169, hep-th/9411048;
Int. J. Mod. Phys. A11 (1996) 1929-1974, hep-th/9505150;   A. Hanany
and Y. Oz, Nucl. Phys. {\bf B452} (1995) 283,
hep-th/9505075;
P.  C.  Argyres, M.  R.  Plesser and A.  D.  Shapere, Phys.  Rev.
Lett.  {\bf 75} (1995) 1699, hep-th/9505100;
P. C. Argyres and A. D. Shapere, Nucl. Phys. {\bf B461} (1996) 437,
hep-th/9509175; 
  A. Hanany, Nucl.Phys. {\bf B466} (1996) 85,  hep-th/9509176.


\bibitem{JR}
R. Jackiw and C. Rebbi, Phys. Rev. {\bf D13} (1976) 3398.


\bibitem{KT}
K. Konishi and H. Terao, Nucl. Phys. {\bf B511} (1998) 264,
hep-th/9707005;   

% G. Carlino, K. Konishi and H. Terao,  JHEP {\bf  04}
%(1998) 003, hep-th/9801027.

\bibitem{Witten} E.~Witten, Phys. Lett. \textbf{86B}, 283  (1979).

\bibitem {TH}  G. 't Hooft, {Nucl. Phys.} {\bf  B190}   (1981) 455.

 \bibitem{APS}   P.C.~Argyres, M.R.~Plesser and N.~Seiberg, Nucl. Phys.  B {\bf 471}, 159  
(1996), hep-th/9603042. 


\bibitem{Others}  P.C.~Argyres, M.R.~Plesser and  A.D.~Shapere, 
Nucl. Phys. B {\bf 483}, 172 (1997), hep-th/9608129;
P.C. Argyres, M.R. Plesser, and A.D. Shapere,
Nucl. Phys.  {\bf B483}   (1997) 172,   hep-th/9608129;
 K.~Hori, H.~Ooguri and Y.~Oz,  Adv. Theor. Math. Phys.   {\bf 1}, 1  (1998), hep-th/9706082.


\bibitem{SYrec}
   M.~Shifman and A.~Yung,
    %``Non-Abelian Duality and Confinement in N=2 Supersymmetric QCD,''
  Phys.\ Rev.\ D {\bf 79}, 125012 (2009), arXiv:0904.1035 [hep-th],
  %%CITATION = ARXIV:0904.1035;%%
    %``Non-Abelian Confinement in N=2 Supersymmetric QCD: Duality and Kinks on Confining Strings,''
  Phys.\ Rev.\ D {\bf 81}, 085009 (2010), arXiv:1002.0322 [hep-th],
  %%CITATION = ARXIV:1002.0322;%%
    %``Non-Abelian Duality and Confinement: from N=2 to N=1 Supersymmetric QCD,''
  Phys.\ Rev.\ D {\bf 83}, 105021 (2011), arXiv:1103.3471 [hep-th],
  %%CITATION = ARXIV:1103.3471;%%
   %``Confronting Seiberg's Duality with $r$ Duality in ${\mathcal N}=1$ Supersymmetric QCD,''
  arXiv:1204.4164 [hep-th],
  %%CITATION = ARXIV:1204.4164;%%
  %``r Duality and 'Instead-of-Confinement' Mechanism in N=1 Supersymmetric QCD,''
  arXiv:1204.4165 [hep-th];  
  %%CITATION = ARXIV:1204.4165;%%
  A.~Marshakov and A.~Yung,
  %``Strong versus Weak Coupling Confinement in N=2 Supersymmetric QCD,''
  Nucl.\ Phys.\ B {\bf 831}, 72 (2010), 
  arXiv:0912.1366 [hep-th].
  %%CITATION = ARXIV:0912.1366;%%

\bibitem{BKM} 
  S.~Bolognesi, K.~Konishi and G.~Marmorini,
  %``Light nonAbelian monopoles and generalized r-vacua in supersymmetric gauge theories,''
  Nucl.\ Phys.\ B {\bf 718}, 134 (2005), hep-th/0502004.
  %%CITATION = HEP-TH/0502004;%%
  
  
  \bibitem{Simone}  
%\cite{Giacomelli:2012ea}
  S.~Giacomelli,
  %``Singular points in N=2 SQCD,''
  JHEP\ 1209 (2012) 040,
  arXiv:1207.4037 [hep-th].
  %%CITATION = ARXIV:1207.4037;%%

\bibitem{noi}
L. Di Pietro and S. Giacomelli,
%``Confining vacua in SQCD, the Konishi anomaly and the Dijkgraaf-Vafa superpotential,''
JHEP\ 1202 (2012) 087, 
arXiv:1108.6049 [hep-th].

  %\cite{Cachazo:2002ry}
\bibitem{CDSW} 
  F.~Cachazo, M.~R.~Douglas, N.~Seiberg and E.~Witten,
  %``Chiral rings and anomalies in supersymmetric gauge theory,''
  JHEP {\bf 0212}, 071 (2002), hep-th/0211170;   
  %%\cite{Cachazo:2002zk}
%\bibitem{Cachazo:2002zk} 
  F.~Cachazo, N.~Seiberg and E.~Witten,
  %``Phases of N=1 supersymmetric gauge theories and matrices,''
  JHEP {\bf 0302}, 042 (2003), hep-th/0301006;
  %%CITATION = HEP-TH/0301006;%%
  F.~Cachazo, N.~Seiberg and E.~Witten,
  %``Chiral rings and phases of supersymmetric gauge theories,''
  JHEP {\bf 0304}, 018 (2003),
  hep-th/0303207.
  %%CITATION = HEP-TH/0303207;%%

%\cite{Gaiotto:2010jf}

%
%\bibitem{Sei}
%N.~Seiberg,   Nucl. Phys. {\bf B435} (1995) 129, hep-th/9411149.


\bibitem{EHIY}
T. Eguchi,  K. Hori, K. Ito and S.-K. Yang, Nucl. Phys. {\bf B471}
(1996) 430, hep-th/9603002.

\bibitem{Bolo} 
  S.~Bolognesi,
  %``A Coincidence Problem: How to Flow from N=2 SQCD to N=1 SQCD,''
  JHEP {\bf 0811}, 029 (2008), arXiv:0807.2456 [hep-th].
  
  
  \bibitem{AMT}
P. C. Argyres, K. Maruyoshi and Y. Tachikawa,
%``Quantum Higgs branches of isolated N=2 superconformal field theories,'' 
arXiv:1206.4700 [hep-th].


\bibitem{ShapTach}
%\cite{Shapere:2008un}
%\bibitem{Shapere:2008un} 
  A.~D.~Shapere and Y.~Tachikawa,
  %``A Counterexample to the 'a-theorem',''
  JHEP {\bf 0812}, 020 (2008), 
  arXiv:0809.3238 [hep-th].
  %%CITATION = ARXIV:0809.3238;%%

\bibitem{NM} Y. Nambu, {Phys. Rev.} {\bf  D10}  (1974) 4262,
   S. Mandelstam, {Phys. Lett.} {\bf 53B}  (1975) 476, 
                               {Phys. Rep.} {\bf 23C} (1976) 245.


\end{thebibliography}
\end{document}